\pdfoutput=1

\documentclass{llncs}

\usepackage[T1]{fontenc} \usepackage[utf8]{inputenc}
\usepackage{url}
\usepackage{graphicx}
\DeclareGraphicsExtensions{.pdf,.png,.jpg}
\graphicspath{{./img/}}
\usepackage{algpseudocode}
\usepackage{flushend}
\usepackage{amsmath}
\usepackage{float}
\usepackage[font=small,labelfont=bf]{caption}
\usepackage{amssymb}
\usepackage{glossaries}
\usepackage[super]{nth}
\usepackage{mathptmx}
\usepackage{todonotes}

\usepackage{tikz}

\usepackage{booktabs}
\usepackage{algpseudocode}
\usepackage{algorithm}
\usepackage{varwidth}
\usepackage{relsize}

\DeclareMathSizes{9}{9}{7.5}{6.5}

\usepackage{multirow}
\usepackage{textcomp}
\usepackage{wrapfig}
\usepackage{lipsum}
\usepackage{dcolumn}
\usepackage{cleveref}

\usepackage{snapshot}
\usepackage{makeidx}
\usepackage{mathtools}
\DeclarePairedDelimiter{\ceil}{\lceil}{\rceil}

\newcommand{\coe}{$CO_{2}e$}

\newcommand{\eqtopmargin}{-0.5cm}
\newcommand{\eqbottommargin}{-0.5cm}
\newcommand{\tabletopmargin}{-0.1cm}

\newcommand{\tablebottommargin}{-0.1cm}
\newcommand{\figtopmargin}{-0.2cm}
\newcommand{\figbottommargin}{-0.5cm}
\newcommand{\figcaptionmargin}{-0.7cm}
\newcommand{\sectionendmargin}{-0.3cm}
\newcommand{\sectiontitlemargin}{-0.3cm}
\newcommand{\subsectionendmargin}{-0.3cm}
\newcommand{\subsectiontitlemargin}{-0.2cm}

\newacronym{coe}{\coe}{$CO_{2}$ equivalent}
\newacronym{ewma}{EWMA}{exponentially weighted moving average}
\newacronym{dvfs}{DVFS}{dynamic voltage \& frequency scaling}
\newacronym{vm}{VM}{virtual machine}
\newacronym{era}{ERA}{energy reduction assets}
\newacronym{api}{API}{application programming interface}
\newacronym{os}{OS}{operating system}
\newacronym{rtp}{RTP}{real-time pricing}
\newacronym{qos}{QoS}{quality of service}
\newacronym{sla}{SLA}{service level agreement}
\newacronym{slo}{SLO}{service level objective}
\newacronym{rtep}{RTEP}{real-time electricity pricing}
\newacronym{iaas}{IaaS}{infrastructure as a service}
\newacronym{paas}{PaaS}{platform as a service}
\newacronym{saas}{SaaS}{software as a service}
\newacronym{pm}{PM}{physical machine}
\newacronym{pue}{PUE}{power usage efficiency}
\newacronym{cue}{CUE}{carbon usage effectiveness}
\newacronym{cef}{CEF}{carbon emission factor}
\newacronym{hpc}{HPC}{high-performance computing}
\newacronym{db}{DB}{database}
\newacronym{dc}{DC}{data center}
\newacronym{oltp}{OLTP}{online transaction processing}
\newacronym{mse}{MSE}{mean squared error}
\newacronym{ga}{GA}{genetic algorithm}
\newacronym{tc}{TC}{treatment category}
\newacronym{wtp}{WTP}{willingness to pay}

\title{Energy-Aware Cloud Management through\\Progressive SLA Specification}

\author{Dražen Lučanin\inst{1} \and Foued Jrad\inst{2} \and Ivona Brandic\inst{1} \and Achim Streit\inst{2}}

\institute{Vienna University of Technology, Austria,
\email{\{drazen,ivona\}@infosys.tuwien.ac.at}
\and Karlsruhe Institute of Technology, Germany,
\email{\{foued.jrad,achim.streit\}@kit.edu}}

\begin{document}

\vspace{-3cm}

\maketitle

\newcommand{\tdown}{400 s}
\newcommand{\revenueincrease}{$43\%$}
\newcommand{\customerincrease}{$5.89\times$}
\newcommand{\ensavings}{$39\%$}

\newcommand{\slanum}{eight}
\newcommand{\slanumnum}{8}
\newcommand{\slanummult}{1--60}
\newcommand{\slasall}{\gls{sla} 1-8}
\newcommand{\slaspp}{\gls{sla} 3-8}
\newcommand{\gadown}{27 minutes}
\newcommand{\gaav}{98.12\%}

\newcommand{\DRmin}{12.5\%}
\newcommand{\DRmax}{66.67\%}

\vspace{-0.7cm}

\begin{abstract}


Novel energy-aware cloud management methods dynamically reallocate computation across geographically distributed data centers to leverage regional electricity price and temperature differences. As a result, a managed \gls{vm} may suffer occasional downtimes. Current cloud providers only offer high availability \gls{vm}s, without enough flexibility to apply such energy-aware management. In this paper we show how to analyse past traces of dynamic cloud management actions based on electricity prices and temperatures to estimate \gls{vm} availability and price values. We propose a novel \gls{sla} specification approach for offering \gls{vm}s with different availability and price values guaranteed over multiple \gls{sla}s to enable flexible energy-aware cloud management. We determine the optimal number of such \gls{sla}s as well as their availability and price guaranteed values. We evaluate our approach in a user \gls{sla} selection simulation using Wikipedia and Grid’5000 workloads. The results show higher customer conversion and \ensavings{} average energy savings per \gls{vm}.

\vspace{-0.3cm}

\end{abstract}

\vspace{-1cm}

\section{Introduction}

\vspace{\sectiontitlemargin}

Energy consumption of data centers accounts for 1.5\% of\
global electricity usage \cite{koomey_worldwide_2008}\
and annual electricity bills\
of \$40M for large cloud providers \cite{qureshi_cutting_2009}.\
In an effort to reduce energy demand,\
new energy-aware cloud management methods leverage geographical data center\
distribution along with location- and time-dependent factors such as\
electricity prices \cite{weron_modeling_2006}\
and cooling efficiency \cite{zhou_optimization_2012}\
that we call \emph{geotemporal inputs}.\
By dynamically reallocating computation based on geotemporal inputs,\
promising cost savings can be achieved \cite{qureshi_cutting_2009}.

Energy-aware cloud management may reduce \gls{vm} availability,\
because certain management actions like \gls{vm} migrations\
cause temporary downtimes \cite{liu_performance_2011}.\
As long as the resulting availability is higher than the value\
guaranteed in the \gls{sla}, cloud providers\
can benefit from the cost savings. However, current cloud providers only offer\
high availability \gls{sla}s, e.g. 99.95\% in case of Google and Amazon.\
Such \gls{sla}s do not leave enough flexibility\
to apply energy-aware cloud management or result in \gls{sla} violations.

Alternative \gls{sla} approaches exist, such as auction-based\
price negotiation in Amazon spot instances \cite{chen_tradeoffs_2011}\
or calculating costs per resource utilisation \cite{berndt_towards_2013}.\
However, estimating availability and price values that can be guaranteed\
in \gls{sla}s for \gls{vm}s managed based on geotemporal inputs\
is still an open research issue. This problem is challenging,\
because exact \gls{vm} availability and energy costs depend\
on electricity markets, weather conditions, application memory access patterns\
and other volatile factors.

In this paper, we propose a novel approach for estimating\
the optimal number of \gls{sla}s, as well as their\
availability and price values under energy-aware cloud management.\
Specifically, we present a method to analyse past traces\
of dynamic cloud management actions based on geotemporal inputs\
to estimate \gls{vm} availability and price values\
that can be guaranteed in an \gls{sla}.\
Furthermore, we propose a progressive \gls{sla} specification\
where a \gls{vm} can belong to one of multiple treatment categories,\
where a treatment category defines the type of energy-aware management actions\
that can be applied. An \gls{sla} is generated for each treatment category\
using our availability and price estimation method.

We evaluate our method by estimating availability and price values\
for \gls{sla}s of \gls{vm}s managed by\
two energy-aware cloud management schedulers --\
the live migration scheduler adapted for clouds\
from \cite{abbasi_dynamic_2011,liu_renewable_2012}\
and the peak pauser scheduler \cite{lucanin_take_2013-1}.\
We evaluate the SLA specification in a user \gls{sla} selection simulation\
based on multi-auction theory \cite{jrad2013}\
using Wikipedia and Grid’5000 workloads to represent multiple user types.\
Our results show that more users with different requirements\
and payment willingness can find a matching \gls{sla} using our specification,\
compared to existing high availability \gls{sla}s.\
Average energy savings of \ensavings{} per \gls{vm} can be achieved\
due to the extra flexibility of lower availability \gls{sla}s.\
Furthermore, we determine the optimal number of offered \gls{sla}s\
based on customer conversion.

\vspace{\sectionendmargin}

\section{Related Work}
\label{sec:related}
\vspace{\sectiontitlemargin}

The related work we cover in this section can be split into two categories:\
(1) energy-aware distributed\
system management methods adapted to geotemporal inputs and\
(2) alternative \gls{vm} pricing models suited for\ 
energy-aware cloud management.\

Qureshi et al. simulate gains from temporally- and geographically-aware\
content delivery networks in \cite{qureshi_cutting_2009},\
with predicted electricity cost\
savings of up to 45\%.\
Lin et al. \cite{lin_online_2012} analyse a scenario with temporal\
variations in electricity prices and renewable energy availability\
for computation consolidation.\
%
Liu et al. \cite{liu_renewable_2012} define an algorithm for power demand\
shifting\ 
according to renewable power availability\
and cooling efficiency.\ 
A job scheduling algorithm for geographically-distributed data centers with\
temperature-dependent cooling efficiency\
is given in \cite{xu_temperature_2013}.\
A method for using migrations across geographically distributed data centers\
based on cooling efficiency is shown in \cite{le_reducing_2011}.\
These approaches, however, do not consider the implications\ 
of energy-aware cloud management on \gls{qos} and costs\
in the \gls{sla} specification.


The disadvantages of current \gls{vm} pricing models\
relying on constant rates\
have been shown by Berndt et al. \cite{berndt_towards_2013}.\
They propose a method for variable \gls{vm} pricing,\
based on the actual \gls{vm} utilisation.\ 
A new charging model for \gls{paas} providers, where variable-time requests\
can be specified by the users,\
is developed in \cite{vieira_towards_2013}.\
Ibrahim et al. applied machine learning\ 
to compensate interferences\
between \gls{vm}s for a pay-as-you-consume\
pricing scheme \cite{ibrahim_towards_2011}.\
Though related, Amazon spot instances \cite{chen_tradeoffs_2011}\
permanently terminate\
\gls{vm}s that get outbidden, hence requiring fault-resilient\
application architectures.\
Aside from this, they\ 
perform exactly like other Amazon instances, \ 
again not allowing temporary downtimes necessary every day\
for energy-aware cloud management.\
Amazon spot instances do show that end users are willing to accept\
a more complex pricing model to lower their costs for certain applications,\
indicating the feasibility of such approaches in real cloud deployments.\
\
None of the mentioned pricing approaches consider\ 
energy-aware cloud management based on geotemporal inputs or\
the accompanying \gls{qos} and energy cost uncertainty,\
which is the focus of our work.

\vspace{\sectionendmargin}
\section{Progressive \gls{sla} Specification}
\vspace{\sectiontitlemargin}
\vspace{-0.4cm}
\label{sec:approach}
\
\



\begin{figure}
\vspace{\figtopmargin}
\vspace{-0.5cm}
\centering
\includegraphics[width=0.98\textwidth]{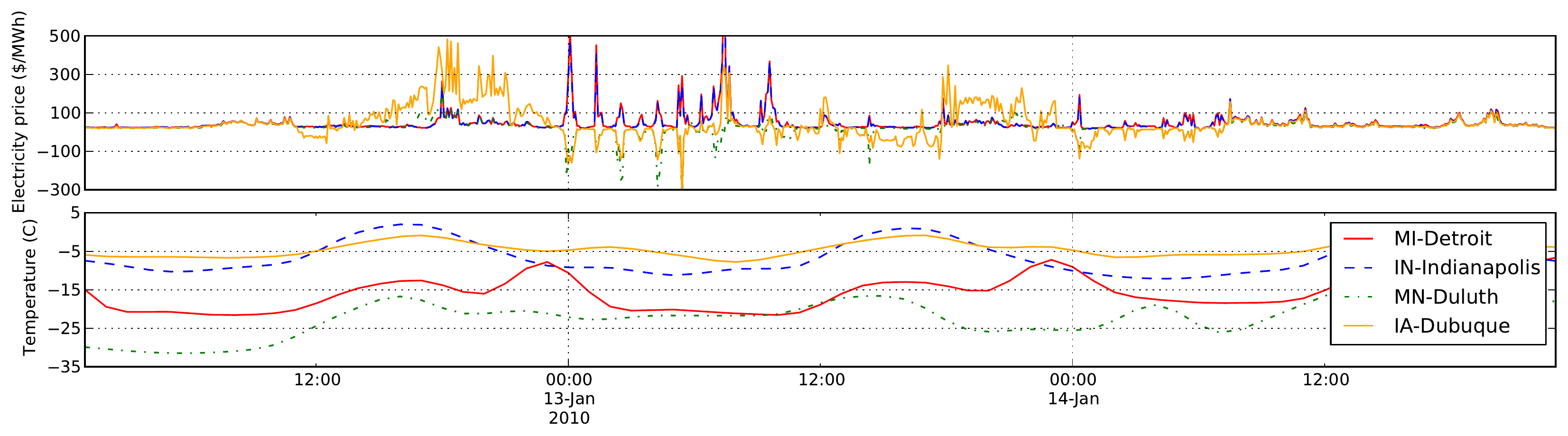}
\vspace{-0.2cm}
\caption{Real-time electricity prices and temperatures in four US cities over three days}
\label{fig:geotemporal_inputs}
\vspace{-0.1cm}
\vspace{\figbottommargin}
\end{figure}

To be able to reason about energy-aware cloud management\
in terms of \gls{sla} specification,\
we analyse two concrete schedulers:\
\
(1) migration scheduler\
(adapted for clouds from \cite{abbasi_dynamic_2011,liu_renewable_2012}) -- \
applies a genetic algorithm to dynamically migrate\
VMs, such that energy costs based on geotemporal inputs are minimised, while\
also minimising the number of migrations per \gls{vm}\
to retain high availability.\
(2) peak pauser scheduler \cite{lucanin_take_2013-1} --\
\
pauses the managed \gls{vm}s for a predefined duration every day,\
choosing the hours of the day that are statistically most likely\
to have the highest energy cost, thus reducing \gls{vm} availability, but also\
the average energy cost.\
Both schedulers depend on geotemporal inputs.
To illustrate the inputs affecting scheduling decisions,\
a three-day graph of real-time electricity prices and temperatures\
in four US cities is shown in Fig.~\ref{fig:geotemporal_inputs}\
(we describe the source dataset in the following section).\
We can see rapid changes in electricity prices on January 13th,\
and very small changes on the following day.\
Subsequently, more frequent \gls{vm} migrations\
would be triggered on the first day,\
in case of the migration scheduler.\
As we cannot predict future geotemporal inputs with 100\% accuracy,\
we can neither exactly predict the scheduling actions that would be applied.\
Instead, only an estimation based on historical data is possible.\

With such energy-aware cloud management methods in mind,\
we propose a progressive \gls{sla} specification,\
where services are divided among multiple treatment categories,\
each under a different \gls{sla} with different availability and price values.\
Hence, different schedulers can be used for \gls{vm}s\
in different treatment categories\
(or the same scheduler with different \gls{qos} constraint parameters).\
The goal of this approach is\
to allow different levels of energy-aware cloud management\
and thus achieve higher energy savings on \gls{vm}s\
with lower availability requirements.\
What is given, therefore, are the schedulers for each treatment category,\
and the historical traces of generated schedules.\
What we have to find are the availability and price values\
that can be guaranteed in the SLAs for each treatment category\
and the optimal number of such SLAs\
to balance SLA flexibility and search difficulty for users.

\begin{figure}
\vspace{\figtopmargin}
\vspace{-0.3cm}
\vspace{-0.2cm}
\centering
\includegraphics[width=0.8\textwidth]{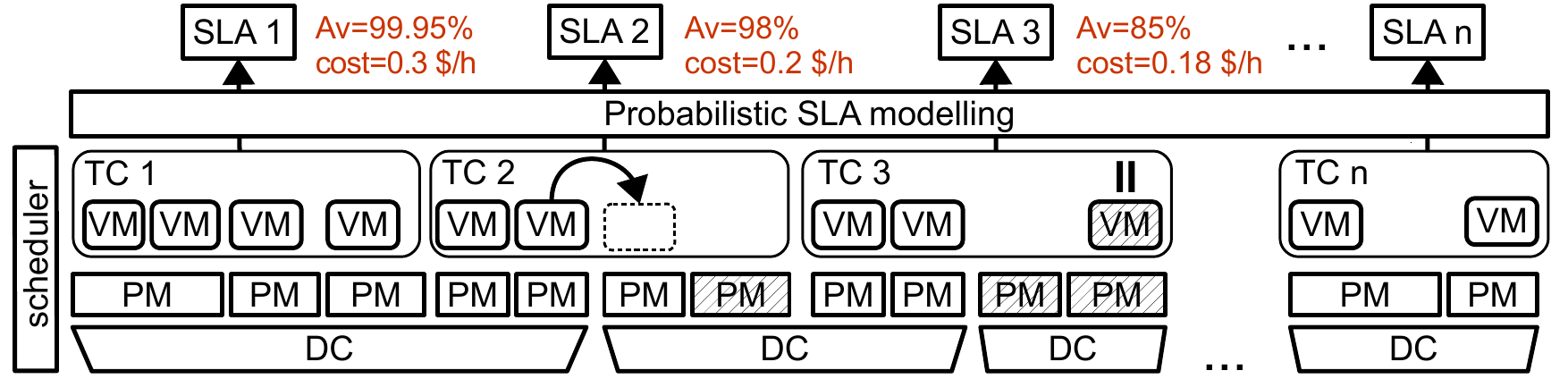}
\caption{Progressive \gls{sla} specification}
\label{fig:vmpricing-overview}
\vspace{-0.2cm}
\vspace{-0.1cm}
\vspace{\figbottommargin}
\end{figure}

We illustrate this approach in Fig.~\ref{fig:vmpricing-overview}.\
A cloud provider operates a number of \gls{vm}s,\
each hosted on a \gls{pm} located\
in one of the geographically-distributed data centers (DC).\
Each \gls{vm} belongs to a \gls{tc} that determines the type of\
scheduling that can be applied to it.\
For example, \gls{tc} 1 is a high availability category where no actions\
are applied on running \gls{vm}s.\
\gls{tc} 2 is a category for moderate cloud management actions, such as\
live \gls{vm} migrations (marked by an arrow) which result in short downtimes.\
\gls{tc} 3 is a more aggressive category where \gls{vm}s can be paused\
(marked as hatched with two vertical lines above it)\
for longer downtimes.\
Other \gls{tc}s can be defined using other scheduling algorithms\
or by varying parameters, e.g. the maximum pause duration.\
The optimal number of \gls{tc}s (and therefore also \gls{sla}s) $n$\
is determined by analysing user\
\gls{sla} selection to have enough variety to satisfy most user types,\
yet not make the search too difficult,\
which we explore in Section~\ref{sec:evaluation}.\
Aside from selecting the number of \gls{sla}s, another task is setting\
availability and price values for every \gls{sla}.\
As energy-aware cloud management that depends on geotemporal inputs\
introduces a degree of randomness into the resulting availability and price\
values of a \gls{vm}, we can only estimate the values that can be guaranteed.\
We do this using a \emph{probabilistic \gls{sla} modelling} method for\
analysing historical cloud management action traces\
to calculate the most likely worst-case availability and average\
energy cost for a \gls{vm} in a \gls{tc}.\
For the example in Fig.~\ref{fig:vmpricing-overview}, sample values\
are given for the \gls{sla}s.\
\gls{sla} 1 might have an availability (Av) of $99.95\%$\
and a high cost of $0.3$ \$/h\
due to no energy-aware management,\
\gls{sla} 2 might have slightly lower values\
due to live migrations being applied on \gls{vm}s in \gls{tc} 2,\
\gls{sla} 3 might have even lower values due to longer downtimes\
caused by \gls{vm} pausing\ldots\ 
In the following section,\
we will show how to actually estimate availability and price\
values for the \gls{sla}s using probabilistic \gls{sla} modelling.\

\vspace{\sectionendmargin}

\section{Probabilistic \gls{sla} Modelling}
\label{sec:pricing_model}
\vspace{\sectiontitlemargin}
To estimate availability and price values\
that can be guaranteed in an \gls{sla} using probabilistic modelling\
for a certain \gls{tc}, we require historical cloud management traces.\
Cloud management traces\
can be obtained through monitoring,\
but for evaluation purposes we simulate different\
scheduling algorithm behaviour.\ 
\
\gls{vm} price is estimated\
by accounting for the average energy costs.\
To calculate \gls{vm} availability,\ 
we analyse the factors that cause \gls{vm} downtime.\
While the downtime duration of the peak pauser scheduler can be specified\
beforehand, the total downtime caused by the migration scheduler depends\
on \gls{vm} migration duration and rate as dictated by geotemporal inputs,\
so we individually analyse both factors.


\vspace{\subsectionendmargin}

\subsection{Cloud Management Simulation}
\label{sec:cloud-simulation}
\vspace{\subsectiontitlemargin}
Our modelling method can be applied to different\
scheduling algorithms and cloud environments.\
To generate a concrete \gls{sla} offering for evaluation purposes,\
we consider a use case\
%
of a cloud consisting of six geographically distributed data centers.\
We use a dataset of electricity prices described in \cite{alfeld_toward_2012}\
and temperatures from\
the Forecast web service\ 
\cite{_forecast_????}.\
We represent a deployment\ 
with world-wide data center distribution\ 
shown in Fig.~\ref{fig:cities}.\
Indianapolis and Detroit were used as US locations.\
Due to limited data availability\footnote{US-only electricity price source\
and a limit of free API requests for temperatures.},\
we modified data for\
Mankato and Duluth to resemble Asian locations and\
Alton and Madison to resemble European locations\
(using  inter-continent differences\
in time zones and annual mean values).\
The effects of the migration and peak pauser scheduler\
are determined in a simulation\
using the Philharmonic\
cloud simulator we developed \cite{drazen_lucanin_philharmonic_2014}.\
%
%
%
The cloud simulation parameters\
are summarised in Table~\ref{tab:simulation-cloud}.\
We illustrate the application of the presented methods with this use\
case as a running example.

\noindent\begin{minipage}{\textwidth}
  \begin{minipage}{0.62\textwidth}
    \centering
    
	\includegraphics[width=1.05\textwidth]{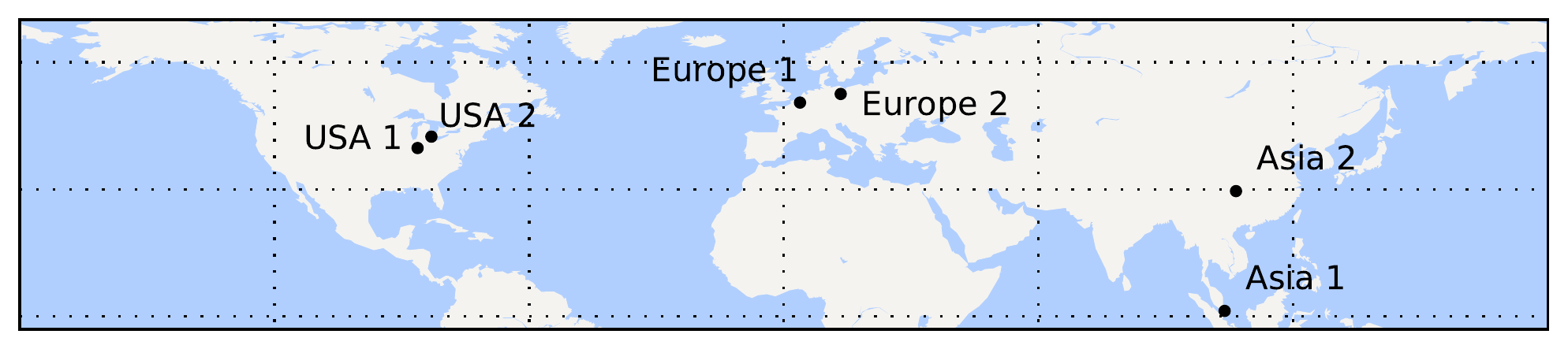}
	\vspace{-0.7cm}
    \captionof{figure}{Simulated world-wide data centers}
    \label{fig:cities}
  \end{minipage}
  \begin{minipage}{0.4\textwidth}
    \centering
	\begin{tabular}{ c c c c }
	\hline
	duration & DCs & PMs & VMs\\
	\hline
	3 months & 6 & 20 & 80 \\
	\hline
	\end{tabular}
	\captionof{table}{Cloud simulation} 
	\label{tab:simulation-cloud}
  \end{minipage}
\end{minipage}



\vspace{\subsectionendmargin}

\subsection{Migration Duration}
\vspace{\subsectiontitlemargin}
Even a live \gls{vm} migration incurs a temporary downtime,\
in the stop-and-copy phase of \gls{vm} memory transferring.\
A very accurate model, with less than 7\% estimation errors,\
for calculating this downtime overhead\
is presented in \cite{liu_performance_2011}.\
The total \gls{vm} downtime during a single live migration $T_{down}$\
is a function of the \gls{vm}'s memory $V_{mem}$, data transmission rate $R$,\
memory dirtying rate $D$, pre-copying termination threshold $V_{thd}$\
and $T_{resume}$, the time necessary to resume a \gls{vm}.

\vspace{\eqtopmargin}
\begin{eqnarray}
T_{down} = \frac{V_{mem} D^n}{R^{n+1}} + T_{resume} \quad \text{, where} \hskip 1em
n = \ceil[\bigg]{ log_{\frac{D}{R}}\frac{V_{thd}}{V_{mem}} }
\end{eqnarray}
\vspace{\eqbottommargin}

\begin{wrapfigure}{r}{0.47\textwidth}
\vspace{-0.2cm}
\vspace{-0.8cm}
\vspace{-0.8cm}
\centering
\includegraphics[width=0.45\textwidth]{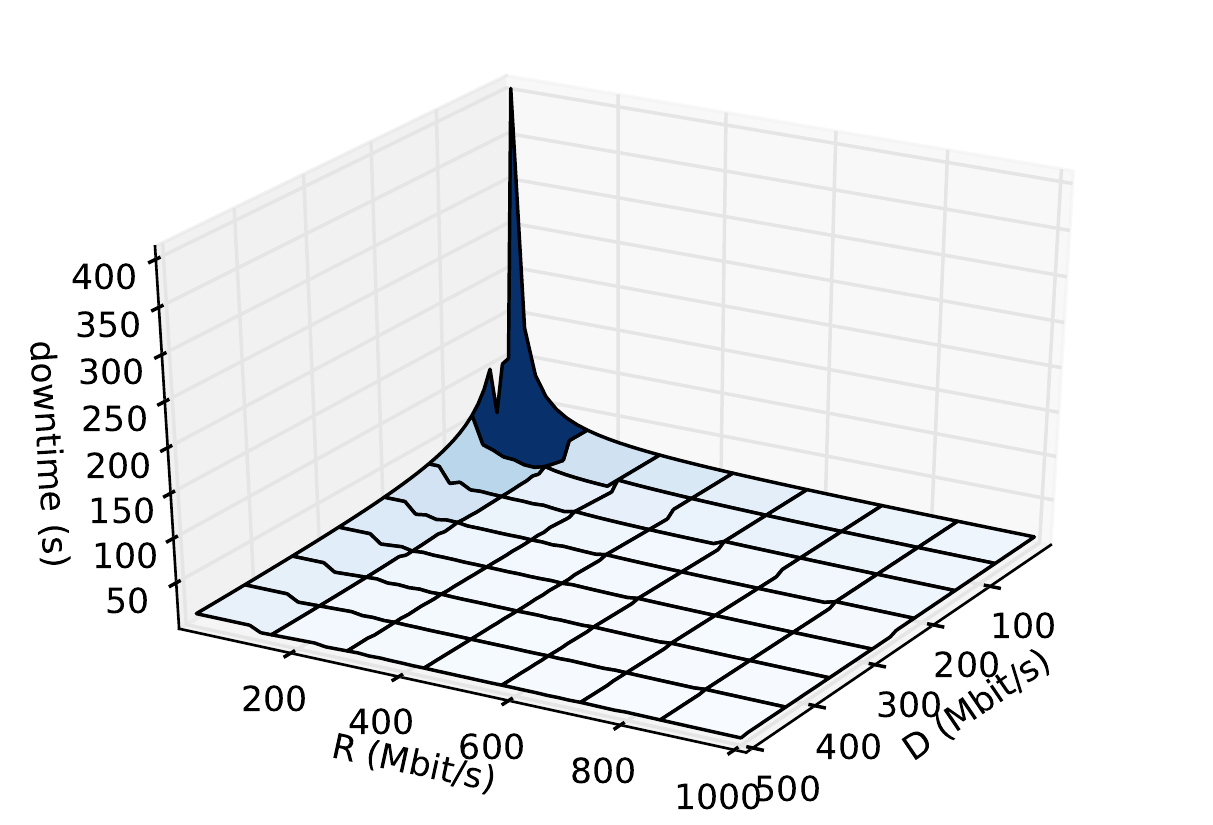}
\caption{\gls{vm} downtime during a live migration}
\label{fig:migration_duration}
\vspace{-0.6cm}
\vspace{-0.6cm}
\end{wrapfigure}

All of the parameters can be determined\
beforehand by the cloud provider, except for $R$ and $D$ which depend\
on the dynamic network conditions and application-specific characteristics.\
Based on historical data, it is possible to reason about the range of these\
variables.\ 
In our running example,\ 
we assume a historical range from low to high values.\
$R$ values in the 10--1000 Mbit/s range were taken based on an\
independent benchmark of Amazon EC2 instance bandwidths.\
$D$ values from 1 kbit/s\
(to represent almost no memory dirtying)\ 
to 1 Gbit/s were taken (the maximum is not important, as will be shown).\
We assumed constant $V_{mem} = 4 GB,\ V_{thd} = 1 GB,\ T_{resume} = 5 s$,\
as these values do not affect the order of magnitude of $T_{down}$.\
We show how $T_{down}$ changes for different $R$ and $D$\ 
in Fig.~\ref{fig:migration_duration}.\
Looking at the graph, we can see that higher $R$ and $D$ values result\
in convergence towards negligible downtime durations and the only area\
of concern is the peak happening when $R$ and $D$ are both very low.\
This happens, because close $R$ and $D$ values lead to a small number\
of pre-copying rounds and copying the whole \gls{vm} under slow speeds\
leads to long downtimes.\
$T_{down}$ is under \tdown{} in the worst-case scenario, \ 
which will be our \gls{sla} estimation as\
suggested in \cite{berndt_towards_2013}.\

\vspace{\subsectionendmargin}

\subsection{Migration Rate}
\vspace{\subsectiontitlemargin}

\begin{figure}
\vspace{\figtopmargin}
\vspace{-0.3cm}
\centering
\begin{minipage}{0.48\textwidth}
    \centering
    \includegraphics[width=\textwidth]{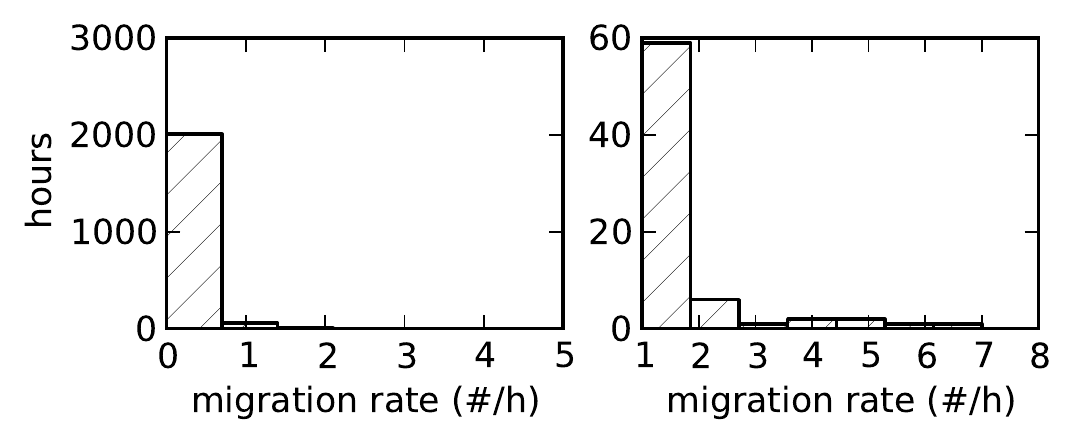}
    \caption{Hourly migration rate histogram}
    \label{fig:migrations-rate}
\end{minipage}
\begin{minipage}{0.48\textwidth}
    \centering
    \includegraphics[width=\textwidth]{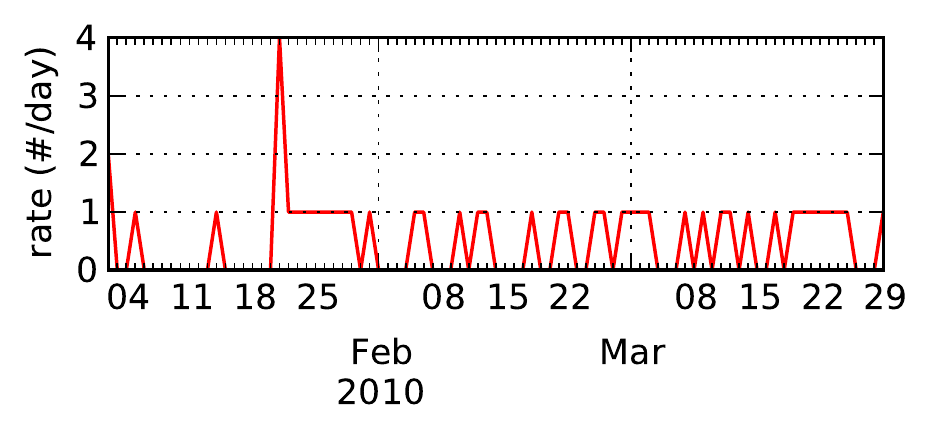}
    \vspace{-0.7cm}
    \caption{Aggregated worst-case migration rate}
    \label{fig:migrations-aggregated}
\end{minipage}
\vspace{\figbottommargin}
\end{figure}

Aside from understanding migration effects, we need to analyse how often they\
occur, i.e. the migration rate.\
We present a method to analyse migration traces\
obtained from the cloud manager's past operation.\
A histogram of migration rates\ 
for the migration traces from our running example\
described in Section~\ref{sec:cloud-simulation}\
can be seen in Fig.~\ref{fig:migrations-rate}.\
The two plots show different zoom levels, as there are few\ 
hours with higher migration rates.\
Most of the time, no migrations are scheduled,\
with one migration per hour happening about 3\% of the time.\

The idea is to group migrations per \gls{vm} and process them in a\
function that aggregates migrations in intervals meaningful to the user.\
We consider an aggregation interval of 24 hours.\
The \textit{aggregated worst-case} function\ 
counts the migrations per \gls{vm} per day and selects\
the highest migration count among all the \gls{vm}s in every interval.\
The output time series is shown in Fig.~\ref{fig:migrations-aggregated}.\
There is one or zero migrations per \gls{vm} most of the time,\
with an occasional case with a higher rate, such as the peak in January.\
Such peaks can occur due to more turbulent geotemporal input changes.\

Given that this data is highly dependent of the scheduling algorithm used\
and the actual environmental parameters,\
fitting one specific statistical distribution to the data to get the desired\
percentile value would be\
hard to generalise for different use cases and might require manual modelling.\
Instead, we propose applying the\
distribution-independent\
bootstrap confidence interval method \cite{efron_introduction_1994}\
to predict the maximum aggregated migration rate.\ 
For our migration dataset,\ 
the 95\% confidence interval for the worst-case migration rate is from three to four\
migrations per day.\

\vspace{\subsectionendmargin}

\subsection{\gls{sla} Options}
\vspace{\subsectiontitlemargin}
\label{sec:8slas}

By combining the migration rate and duration analyses,\
we can estimate the upper bound\
for the total \gls{vm} downtime and, therefore, the availability\
that can be warranted in the \gls{sla}.\
We define availability ($Av$) of a \gls{vm} as:
\
\vspace{-0.1cm}
\begin{equation}
Av = 1 - \frac{\text{total VM downtime}}{\text{total VM lease time}}
\end{equation}
\vspace{-0.05cm}
\
For our migration dataset and the previously discussed migration duration and rate,\
we estimate the total downtime of a \gls{vm}\
controlled by the migration scheduler to be \gadown{} per day\
in the worst case, meaning we can guarantee an availability of \gaav{}.\
We can precisely control the availability of the \gls{vm}s\
managed by the peak pauser.\

The average energy savings ($en\_savings$) for a \gls{vm} running in\
a treatment category $TC_i$ can be calculated by comparing it to the\
high availability $TC_1$.\
From the already described simulation,\ 
we calculate $en\_cost$, the average cost of energy consumed by a \gls{vm}\
based on real-time electricity prices and temperatures.\
We divide the energy costs equally among \gls{vm}s within a \gls{tc}.\
This is an approximation, but serves as an estimation of the\
energy saving differences between \gls{tc}s.\
We calculate energy savings as:\
\
\begin{equation}
en\_savings(VM_{TC_i}) = 1 - \frac{en\_cost(VM_{TC_i})}{en\_cost(VM_{TC_1})}
\label{eq:energy_model}
\end{equation}
\
where $en\_cost(VM_{TC_1})$ is the average energy cost\
for a \gls{vm} in $TC_1$\ with no actions applied\
and $en\_cost(VM_{TC_i})$ is the average energy cost\
for a \gls{vm} in the target $TC_i$.\

The \gls{vm} cost consists of several components.\
\
Aside from $en\_cost$,
$service\_cost$ groups other \gls{vm} upkeep costs\
(manpower, hardware amortization, profit margin etc.)\
during a charge unit (typically one hour in \gls{iaas} clouds).\
We assume the service component to be charged\
linearly to the \gls{vm}'s availability.\
\
\begin{equation}
cost = en\_cost + Av \cdot service\_cost
\label{eq:vm_cost}
\end{equation}
\vspace{\eqbottommargin}
\
To generate the complete \gls{sla} offering\
we consider an Amazon m3.xlarge instance which costs 0.280 \$/h\
(Table~\ref{tab:base-vm}) as a base \gls{vm} with no energy-aware scheduling.\
Base instances with different resource values (e.g. RAM, number of cores)\
can be used, but this is orthogonal to the \gls{qos} requirements of\
availability that we consider and\
would not influence the energy-aware cloud management potential.\
Similarly, Amazon spot instances were not considered specially as they perform\
exactly the same as normal instances while running,\
as we explained in Section~\ref{sec:related}.\
We assumed the service component to be 0.1 \$/h,\
about a third of the \gls{vm}'s price.\
The prices of \gls{vm}s controlled by the two\
energy-efficient schedulers were derived from it,\
applying $en\_savings$ obtained in the cloud simulation.\
The resulting \gls{sla}s are shown in Fig.~\ref{fig:sla}.\ 
SLA 1 is the base \gls{vm}.\ 
SLA 2 is the \gls{vm} controlled 
by the migration scheduler.\
The remaining \gls{sla}s are \gls{vm}s controlled by the peak pauser\
scheduler with downtimes uniformly distributed from \DRmin{} to \DRmax{}\
to represent a wide spectrum of options.\
We chose \slanum{} \gls{sla}s to analyse how \gls{sla} selection changes\
from the user perspective.\
We later show that this number is in the 95\% confidence interval\
for being the optimal number of \gls{sla}s based on our simulation.\
We analyse a wider range of \slanummult{} offered \gls{sla}s\
and how they impact customer conversion\
from the cloud provider's perspective\
in Section~\ref{sec:evaluation}.\
The lines in the background\ 
illustrate value progression.\
$Av$ decreases only slightly for the migration \gls{sla}, yet the energy\
savings are significant due to\
dynamic \gls{vm} consolidation and \gls{pm} suspension.\
For peak pauser \gls{sla}s, availability and costs decrease linearly,\
from high to low values.\ 
The $en\_savings$ values are at first lower\
than those attainable with the migration scheduler,\
as the peak pauser scheduler cannot migrate \gls{vm}s\
to a fewer number of \gls{pm}s, but can only pause them for a certain time.\
With lower availability requirements, however,\
the peak pauser can achieve higher $en\_savings$ and lower prices,\
which could not be reached by \gls{vm} migrations alone.

%
%
%


\noindent\begin{minipage}{\textwidth}
  \begin{minipage}{0.78\textwidth}
    \centering

	\includegraphics[width=1.08\textwidth]{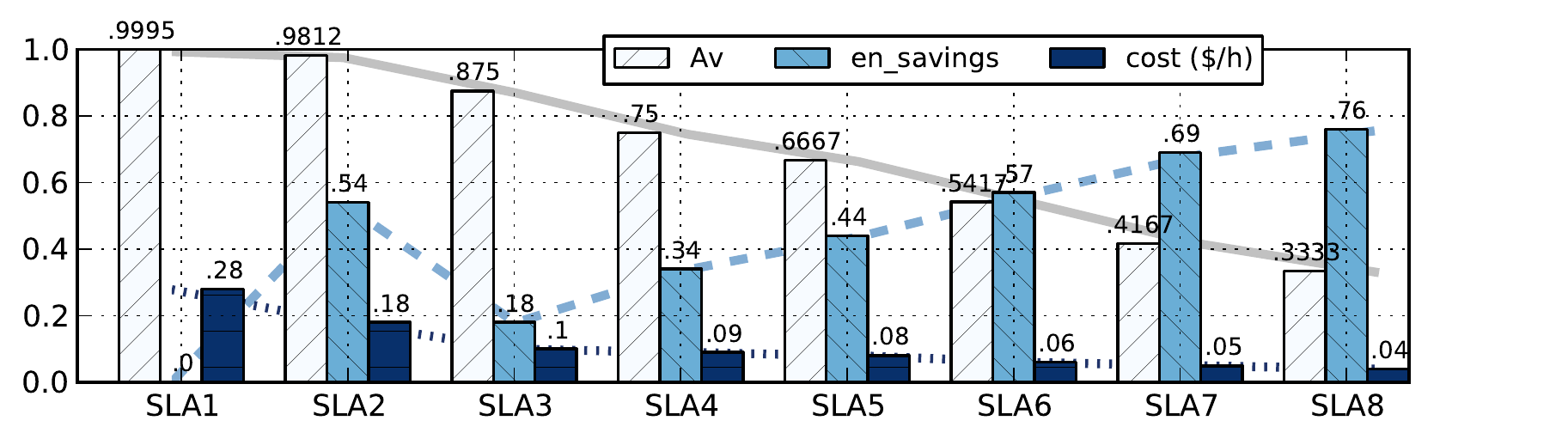}
	\vspace{\figcaptionmargin}
	\captionof{figure}{\gls{sla}s generated for different \gls{tc}s}
	\label{fig:sla}
  \end{minipage}
  \begin{minipage}{0.20\textwidth}
    \centering
	\begin{tabular}{ c c c }
	\hline
	type & m3.xlarge\\
	\hline
	Av & 99.95\%\\
	\hline
	cost & 0.28 \$/h\\
	\hline
	service & 0.1 \$/h\\
	\hline
	\end{tabular}
    \captionof{table}{Base \gls{vm}} 
	\label{tab:base-vm}
  \end{minipage}
\end{minipage}

\vspace{\sectionendmargin}
\vspace{-0.1cm}

\section{User Modelling}
\label{sec:user_model}
\vspace{\sectiontitlemargin}
Knowing the \gls{sla} offering, the next step is\
to model user \gls{sla} selection\
in order to analyse the benefits of our progressive \gls{sla} specification.\
We first describe how we derive user requirements and then the utility\
model used to simulate user \gls{sla} selection based on their requirements.


\vspace{\subsectionendmargin}

\subsection{User Requirements Model}
\vspace{\subsectiontitlemargin}

\begin{figure}
\vspace{\figtopmargin}
\vspace{-0.4cm}
\centering
    \begin{minipage}{0.48\textwidth}
        \centering
        \includegraphics[width=\textwidth]{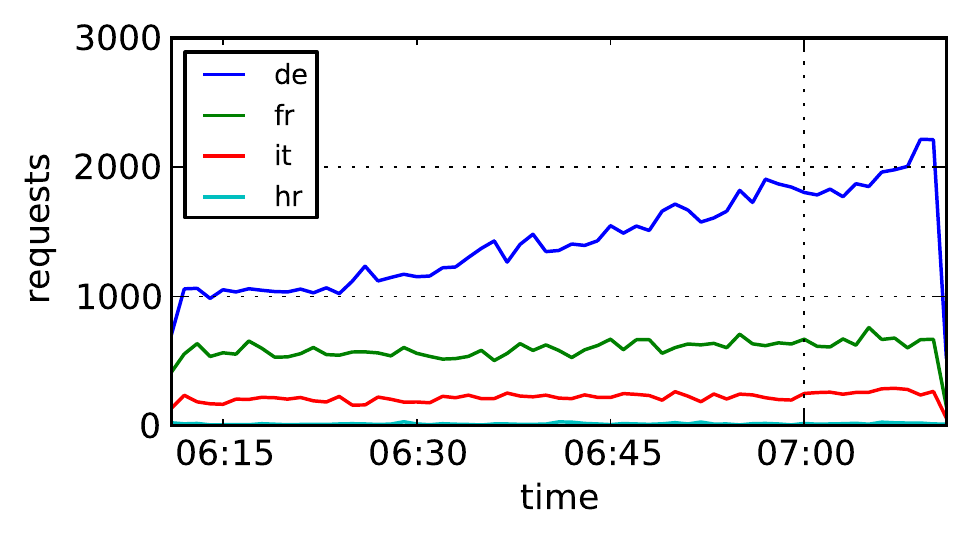}
        \smaller{(a) Wikipedia requests per minute}
    \end{minipage}
    \begin{minipage}{0.48\textwidth}
        \centering
        \includegraphics[width=\textwidth]{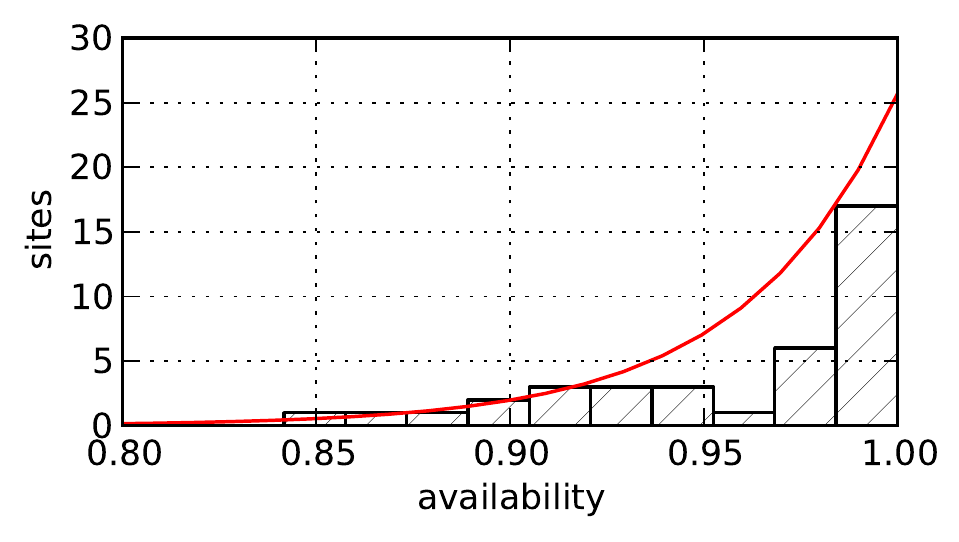}
        \smaller{(b) Wikipedia availability distribution}
    \end{minipage}
\caption{Web user modelling}
\label{fig:user-modelling-web}
\vspace{\figbottommargin}
\end{figure}

To model user requirements, we use real traces\
of web and \gls{hpc} workload,\
since I/O-bound web and CPU-bound \gls{hpc} applications\
represent two major usage patterns of cloud computing.\
As we do not have data on availability requirements of website owners,\
we generate this dataset based on the frequency of end user HTTP requests\
directed at different websites and counting missed requests,\
similarly to how reliability is determined\
from the mean time between failures \cite{oconnor_practical_2011}.\
A public dataset of HTTP requests\
made to Wikipedia \cite{urdaneta_wikipedia_2009} is used.\
To obtain data for different websites, we consider Wikipedia\
in each language as an individual website,\
because of its unique group of end users\
(different in number and usage pattern).\
In this scenario we consider a website owner\
to be the user of an \gls{iaas} service\
(not to be confused with the end user, a website visitor).\
The number of HTTP requests for a small subset of four websites\
(German, French, Italian and Croatian Wikipedia\
denoted by their two-letter country codes)\
is visualised in Fig.~\ref{fig:user-modelling-web} (a)\
for illustration purposes\
(we use the whole dataset with 38 websites for actual requirements modelling).\
The data exemplifies\ 
significant differences in amplitudes.\
Users of the German Wikipedia send between 1k and 2k requests per minute,\
while the Italian and Croatian Wikipedia have less than 300 requests per minute.\
Due to this variability, we assume that different Wikipedia websites\
represent diverse requirements of website owners.\
We model availability requirements by applying a heuristic\
 -- a website's required availability is the minimum\
necessary to keep the number of missed requests\
below a constant threshold (we assume 100 requests per hour).\
Using this heuristic, we built an availability requirements dataset\
for the web user type from 5.6 million requests\
divided among 38 Wikipedia language subdomains.\
The resulting availability requirement histogram\
can be seen in Fig.~\ref{fig:user-modelling-web} (b). It follows an\
exponential distribution (marked in red).\
There is a high concentration of sites that need almost full availability,\
with a long tail of sites that need less (0.85--1.0).

\begin{figure}
\vspace{\figtopmargin}
\vspace{-0.5cm}
\centering
    \begin{minipage}{0.48\textwidth}
        \centering
        \includegraphics[width=\textwidth]{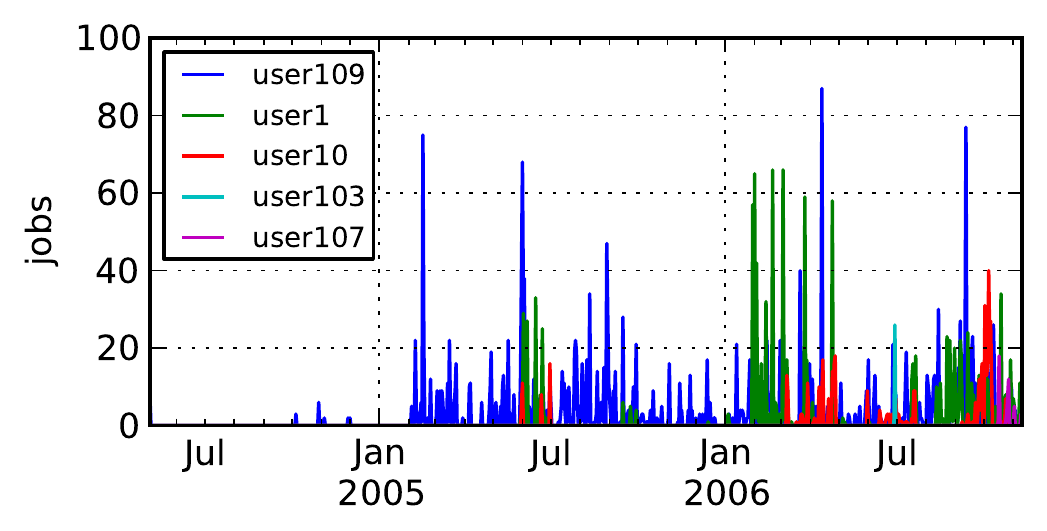}
        \smaller{(a) G5k example job submissions}
    \end{minipage}
    \begin{minipage}{0.48\textwidth}
        \centering
        \includegraphics[width=\textwidth]{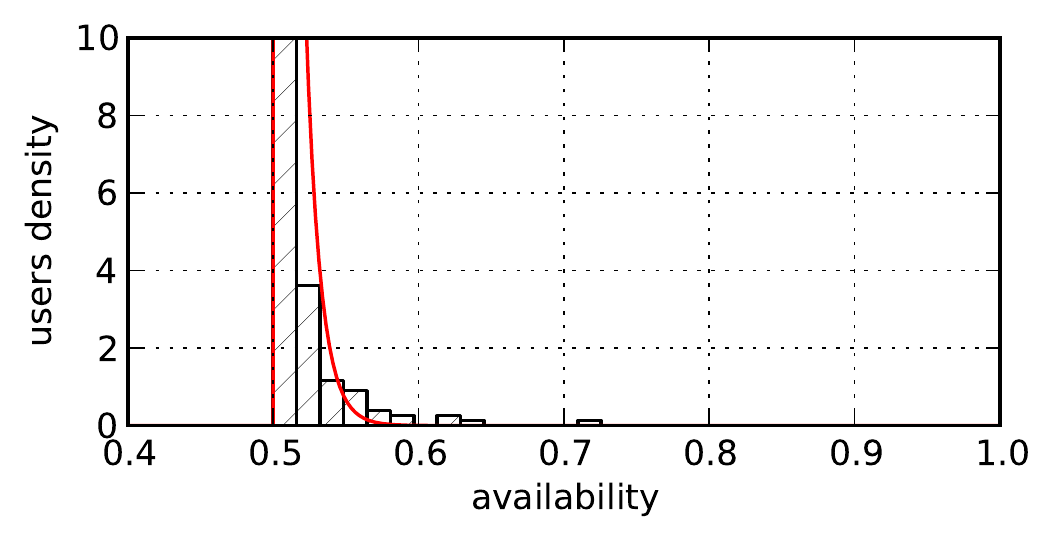}
        \smaller{(b) G5k availability distribution}
    \end{minipage}
\vspace{-0.3cm}
\caption{HPC user modelling}
\label{fig:user-modelling-hpc}
\vspace{-0.2cm}
\vspace{\figbottommargin}
\end{figure}

For HPC workload, we use\
a dataset\ 
of job submissions made to Grid'5000 (G5k)\
\cite{franck_cappello_gwa-t-2_????},\
a distributed job submission platform spread across 9 locations in France.\
The number of jobs submitted by a small subset of users\
is visualised in Fig.~\ref{fig:user-modelling-hpc} (a).\
While some users submit jobs over a wide period (blue), others\
only submit jobs in small bursts (green), but the load is not nearly\
as constant as the web requests from the Wikipedia trace.\
To model HPC users' availability requirements, where jobs have\
variable duration as well as rate (unlike web requests, which typically have\
a very short duration), we use another heuristic.\
Every user's availability requirement is\
mapped between a constant minimum availability (we assume 0.5)\
and full availability using $mean\_duration \cdot mean\_rate$,\
which stands for mean job duration and mean job submission rate per user.\
Using this heuristic, we built a dataset of availability requirements\
for the \gls{hpc} user type from jobs submitted\
over 2.5 years by 481 G5k users.\
The resulting availability requirement distribution\
(normalised such that the area is 1)\
can be seen in Fig.~\ref{fig:user-modelling-hpc} (b).\
The distribution marked red again follows an exponential distribution\
(the first bin, cut off due to the zoom level, shows a density of 100),\ 
but with the tail facing the opposite direction\
than the web requirements.\
\gls{hpc} users submit smaller and less frequent jobs most of the time,\
with a long tail of longer and/or more frequent jobs (from 0.5 to 0.75).\

%
%

Every user's \gls{wtp} is derived by multiplying\
his/her availability requirement with\
the base \gls{vm} price and adding Gaussian noise $\mathcal{N}(0,\,0.05^2)$\
to express subjective value perception.\
We selected the noise standard deviation\
to get positive \gls{wtp} values considering the availability model.\
The resulting \gls{wtp} histogram is shown in Fig.~\ref{fig:wtp}.\
It can be seen that
\gls{hpc} users have lower \gls{wtp} values,
but there is also an overlap area with web users who have similar requirements.


\vspace{\subsectionendmargin}

\subsection{Utility Model}
\vspace{\subsectiontitlemargin}

The utility-based model is used to simulate how users select services\
based on their requirements.\
We use a quasi-linear utility function adopted from\
multi-attribute auction theory \cite{jrad2013}\ 
to quantify the user's preference for a provided \gls{sla}.\
The utility is calculated by multiplying the user's SLA satisfaction score\
with \gls{wtp}\ 
and subtracting the \gls{vm} cost\
charged by the provider.\ 
The utility for user $i$ from selecting a VM instance type $t$\
with availability $Av_t$ is calculated as:
\vspace{-0.1cm}
\begin{equation}
	U_{i}(VM_t)=WTP_i \cdot f_i(Av_t)-cost(VM_t)
	\label{eq:utility}
\end{equation}
\vspace{\eqbottommargin}

\begin{figure}
\vspace{\figtopmargin}
\vspace{-0.4cm}

    \begin{minipage}{0.48\textwidth}
        \centering
        \includegraphics[width=\textwidth]{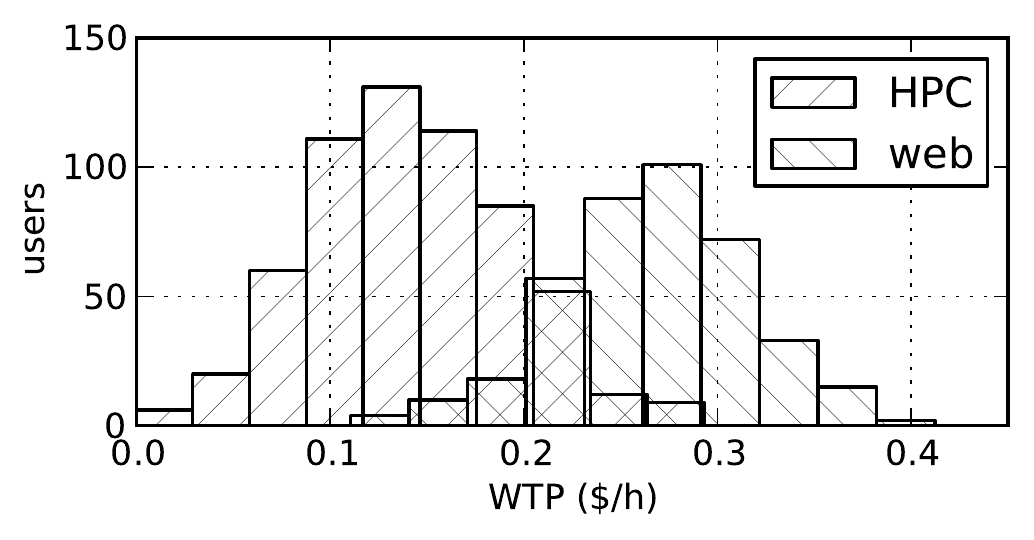}
        \vspace{-0.2cm}
        \caption{\gls{wtp} histogram}
        \label{fig:wtp}
    \end{minipage}
    \begin{minipage}{0.48\textwidth}
        \centering
        \includegraphics[width=\textwidth]{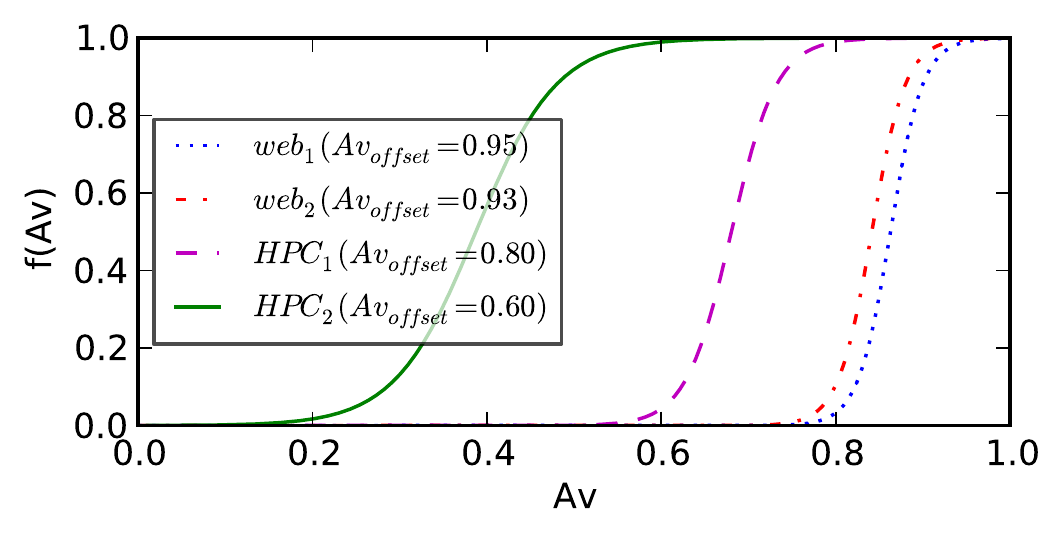}
        \vspace{-0.2cm}
        \caption{SLA satisfaction function}
        \label{fig:satisfaction}
    \end{minipage}

\vspace{-0.2cm}
\vspace{\figbottommargin}
\end{figure}

where 
$f_i(Av_t)$ is the user's satisfaction\
with the offered \gls{vm}'s availability.\
We model it using a mapping function $f_i: [0, 1] \to [0, 1]$,\ 
extended from \cite{jrad2013} with variable slopes:
\
\vspace{-0.15cm}
\begin{equation}
f_{i}(Av_t)=\frac{\gamma}{\gamma+\beta e^{Av_{offset_i}^2 \alpha(Av_{offset_i}-Av_t)}} 
\label{eq:fitting1}
\end{equation}
\
where $Av_{offset_i}$ is the required availability specific to each user,\
which we select based on the exponential models of Wikipedia and G5k data\
presented earlier.\
$\alpha$, $\beta$ and $\gamma$ are positive constants common for all users,\
which we set to 60, 0.01 and 0.99 respectively.\
These values were chosen for a satisfaction of close to\
1 for the desired availability value\
and a steep descent towards 0 for lower values,\
similar to earlier applications of this satisfaction function \cite{jrad2013}.\
The slope of the function also depends on $Av_{offset_i}$, to model that\
users who require a lower availability have a wider range of acceptable values.\
The mapping function is visualised in Fig.~\ref{fig:satisfaction}\
for a small sample of two \gls{hpc} and two web users.\
We can see that for the two web users,\
the slope is almost the same and very steep\
(at 0.93 their satisfaction is close to 1 and at 0.8 it is almost 0).\
The $HPC_1$ user is similar to the web users,\
only with lower availability requirements and a slightly wider slope.\
The $HPC_2$ user has low requirements and a wide slope\
from an availability of 0.2 to 0.6.\
Later in our evaluation, we generate 1000 users,\
each with different mapping functions, distributed\
according to the user requirements model.\

\
User $i$ chooses a VM instance type $VM_{selected}$\
offering the best utility value:\
\
\vspace{-0.1cm}
\begin{equation}
U_{i}(VM_{selected})=\max_{\forall t}\ U_{i}(VM_t)
\end{equation}
\vspace{-0.05cm}
\
unless all types result in a negative utility,\
in which case the user selects none.\
Additionally, we model search difficulty by defining $P_{stop}$, a probability\
that a user will give up the search after an \gls{sla} has been examined.\
We model this probability as increasing after every new \gls{sla} check,\
by having $P_{stop_j} = j \cdot check\_cost$,\
where $j$ is the number of checks already performed\
and $check\_cost$ is a constant parameter standing for the probability\
of stopping after the first check.\
Based on every user's requirements and the \gls{sla} offering, $min\_checks$\
is the minimum number of\
checks necessary to reach a \gls{vm} type that yields a positive $U_i(VM)$.\
We define $P_{quit}$ to be the total probability that\
a user will quit the search before reaching a positive-utility \gls{sla}.
By applying the chain probability rule,\
we can calculate $P_{quit}$ as:\
\
\vspace{-0.1cm}
\begin{equation}\label{eq:pquit}
P_{quit}=\mathlarger{\sum}_{j=1}^{min\_checks - 1} P_{stop_j} \prod_{k=1}^{j-1}(1 - P_{stop_k})
\end{equation}
\vspace{-0.1cm}
\
The outer sum is the joint distribution of all possible\
stop events that may occur for a user\
and the inner product stands for all event outcomes\
when searching continued\
until the $j$-th event was realised as stopping.\ 
We use this expression in the evaluation as a measure of difficulty for\
users to find a matching \gls{sla}.

\vspace{\sectionendmargin}

\section{Evaluation}
\label{sec:evaluation}
\vspace{\sectiontitlemargin}
In this section we describe the simulation of the proposed\
progressive \gls{sla} specification\
using user models based on real data traces\
and analyse the results.


\vspace{\subsectionendmargin}

\subsection{Simulation Environment}
\vspace{\subsectiontitlemargin}

The simulation parameters are summarised in Table~\ref{tab:simulation}.\
The first step of the simulation is to generate\
a population of web and \gls{hpc} users based on the\
requirement models derived from\
the Wikipedia and Grid'5000 datasets, respectively.\
We simulated 1000 users to represent a population with enough variety to\
explore different \gls{wtp} and availability requirements.\
We assume the ratio between web and \gls{hpc} users of 1 : 1.5,\
based on an anlysis of a real system performed in \cite{liu_renewable_2012}.\
We determine each user's \gls{wtp} from the desired availability\
with Gaussian noise, as already explained in the previous section.\
The \gls{sla} offering was derived from the migration and peak pauser\
scheduler using the probabilistic modelling technique\
(Section~\ref{sec:pricing_model}).\
For the examination of \gls{sla} selection from the user's perspective,\
the \slanum{} \gls{sla}s\
we already defined in Section~\ref{sec:8slas} were used.\
To examine the cloud provider's perspective,\
we evaluated \slanummult{} \gls{sla}s,\
doing 100 simulation runs per offering\
to calculate the most likely optimal number of \gls{sla}s.\
A $check\_cost$ of 0.015 is selected to initially start with a low chance of\
the user quitting and then subsequently increase it for every \gls{sla} check\
per Eq.~\ref{eq:pquit}.\
The same $\alpha,\ \beta,\ \gamma$ values that we already explained\
in the previous section were set that result in\
a utility of 1 for the required availability\
and a gradual decline towards a utility of 0 for lower availabilities.\
The core of the simulation is to determine each user's\
\gls{sla} selection (if any) based\
on the utility model (Section~\ref{sec:user_model}).\

\begin{table}
\vspace{\tabletopmargin}
\centering
\caption{Simulation settings}
\label{tab:simulation} 

\begin{tabular}{ l | c c c c c c c c c c}
\hline
\textbf{Parameter} & users & web : \gls{hpc} & \#SLAs \ 
& runs & $check\_cost$ & $\alpha$ & $\beta$ & $\gamma$\\
\hline
\textbf{Value} & 1000 & 1 : 1.5 & \slanummult{} \ 
& 100 & 0.015 & 60 & 0.01 & 0.99\\
\hline
\end{tabular}

\vspace{\tablebottommargin}
\end{table}

\vspace{\subsectionendmargin}

\subsection{User Benefits}
\vspace{\subsectiontitlemargin}

\begin{figure}
\vspace{\figtopmargin}
\vspace{-0.1cm}
\centering
\begin{minipage}{0.48\textwidth}
    \centering
    \includegraphics[width=\textwidth]{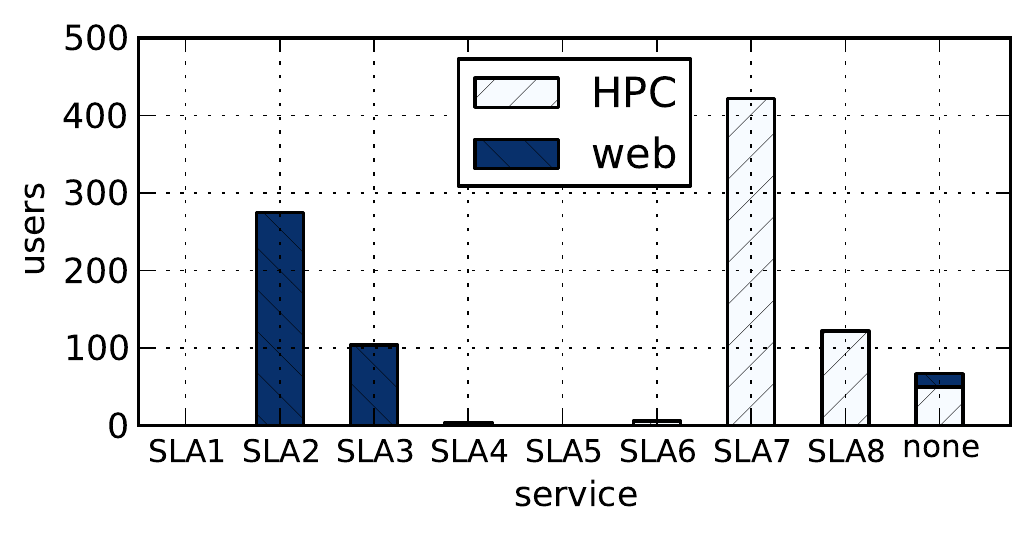}
    \vspace{\figcaptionmargin}
    \caption{Simulated service selection}
    \label{fig:category_distribution}
\end{minipage}
\begin{minipage}{0.48\textwidth}
    \centering
    \includegraphics[width=\textwidth]{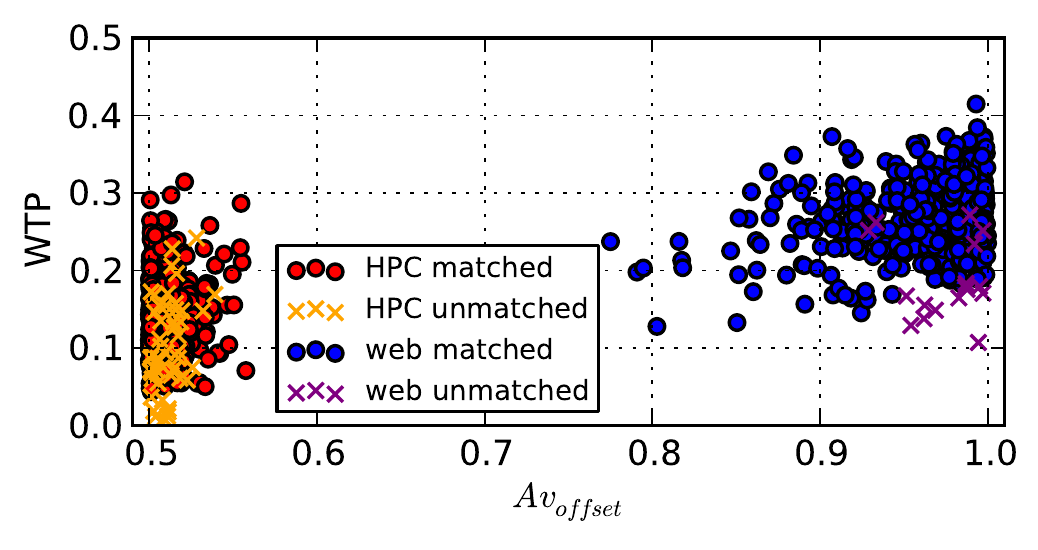}
    \vspace{\figcaptionmargin}
    \caption{Matched and unmatched users}
    \label{fig:unmatched_distribution}
\end{minipage}
\vspace{\figbottommargin}
\end{figure}

Simulation results showing the distribution of users among the\
\slanum{} offered \gls{sla}s from Section~\ref{sec:8slas}\
are presented in Fig.~\ref{fig:category_distribution}.\
Different colours are used for web and HPC users types.\
It can be seen that most of the users successfully found a service\
that matches their requirements, with less than 5\% of unmatched requests.\
The majority of HPC users are distributed between SLA 7 and 8 offering\
42\% and 33\% availability, respectively.\
The majority of web users selected SLA 2, the migration scheduler \gls{tc},\
due to its high availability comparable to a full availability\
service, but a more affordable price due to the energy cost savings.\
26\% of web users opted for SLA 3, the\
peak pauser instance which still offers a high availability (87.5\%),\
but at almost half the price of SLA 2.

The distribution of unmatched users\
who did not select any of the offered services\
(where utility was negative for all \gls{sla}s)\
is shown alongside the matched users\
in Fig.~\ref{fig:unmatched_distribution},\
showing their $Av_{offset}$ and \gls{wtp} values.\
We can see that unmatched users have low \gls{wtp} values,\
the cause of them not being able to find a suitable\
service option.

\vspace{\subsectionendmargin}

\subsection{Cloud Provider Benefits}
\begin{figure}
\vspace{\figtopmargin}
\vspace{-0.1cm}
\centering
\begin{minipage}{0.48\textwidth}
    \centering
    \includegraphics[width=0.9\textwidth]{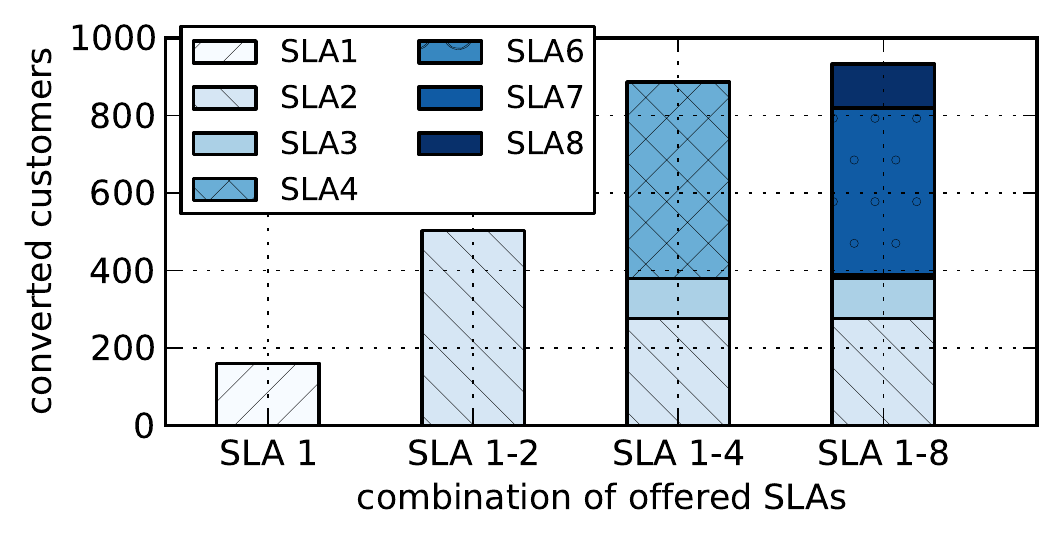}
    \vspace{-0.1cm}
    \caption{Matched users per \gls{sla} combination}
    \label{fig:users_count}
\end{minipage}
\begin{minipage}{0.48\textwidth}
    \centering
    \includegraphics[width=\textwidth]{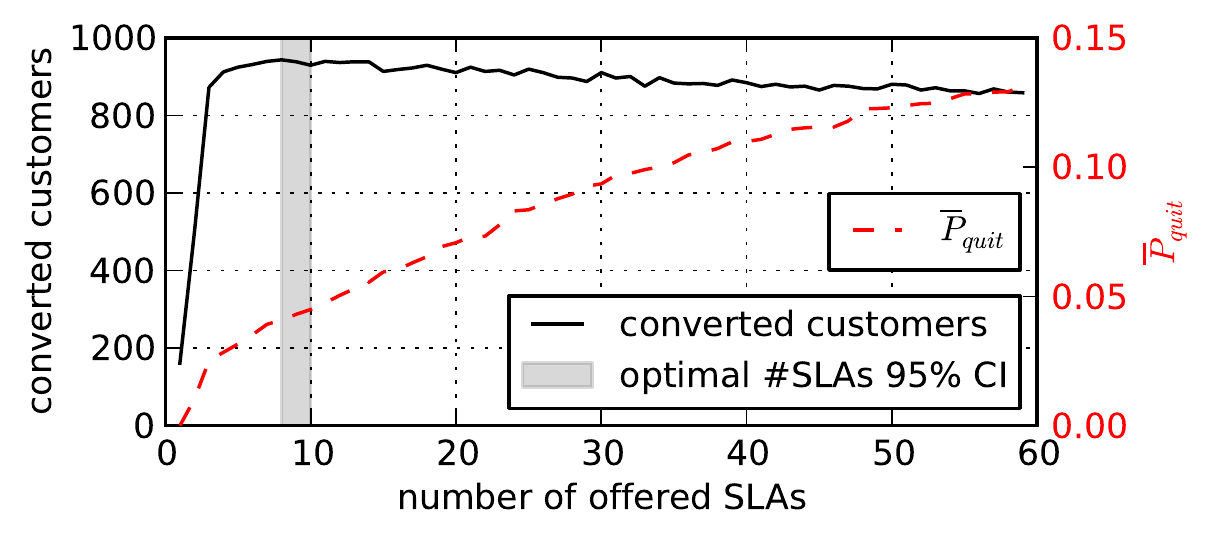}
    \vspace{-0.3cm}
    \caption{Users and $\overline{P}_{quit}$ per \gls{sla} number}
    \label{fig:sla_count}
\end{minipage}
\vspace{\figbottommargin}
\end{figure}

\vspace{\subsectiontitlemargin}
Customer conversion means the number of users\
who looked at the \gls{sla} offering and\
found an \gls{sla} that matches their needs.\
This metric is an indicator of the provider's economic success.\ 
\
To compare the multiple treatment category system with the traditional\
way of only having a full availability option,\
we simulated different \gls{sla} offerings.\
\
Fig.~\ref{fig:users_count} shows customer conversion\ 
with colour indicating the selection distribution\
for different offering combinations\
of the \slanum{} previously examined \gls{sla}s.\
Customer conversion growth can be seen with more service types,\
due to users having\
a higher chance of finding a category that matches\
their requirements.\
\
With \gls{sla}s 1--2 offered, only \gls{sla} 2 was selected,\
as it still offers a high-enough availability to satisfy user requirements\
and the price is lower than in \gls{sla} 1.\
As we widen the offering, more \gls{sla}s get selected, but the majority of\
users choose among two \gls{sla}s that best suit\
the two user types that we modelled.\
Still, a small number of users select other \gls{sla}s\
(\gls{sla} 3 and, if offered, \gls{sla} 8)\
which better suit their needs.\ 
\gls{sla} 5 is never selected\
due to user requirements\
and in real clouds such \gls{sla}s should be removed\
to simplify selection.

The introduced service types can be managed in a more energy efficient\
manner. The average energy savings weighted based on the lease time\
per \gls{vm} for the \slasall{} offering,\
compared to the current 99.95\% availability Amazon instances represented by\
\gls{sla} 1, are \ensavings{}.\
Full annual lease time was assumed for web users\
(as web applications are typically running all the time)\
and was varied based on job runtime and frequency for \gls{hpc} users\
(we assume that a \gls{vm} is provisioned just to perform the submitted job).\
This shows that more energy efficient management is possible if users\
declare the \gls{qos} levels they require through \gls{sla} selection.
For the \slasall{} scenario, where \customerincrease{} more users\
can be converted and the annual lease times explained above,\
a \revenueincrease{} revenue increase is calculated\
from the service component of the selected \gls{vm}s.\
Exact numbers\ 
depend on the user type ratio\
that will vary between cloud providers.\

To find the optimal number of offered \gls{sla}s,\
we performed a simulation where we explore customer conversion\
for a higher number of \gls{sla}s.\
The extra \gls{sla}s were generated for the peak pauser scheduler,\
which allows for arbitrary control of \gls{vm} availability and price.\
The peak pauser \gls{sla}s were uniformly interpolated\
between full and no availability\
to avoid duplicates.\
Fig.~\ref{fig:sla_count} shows how the number of offered \gls{sla}s affects\
the user conversion count and $\overline{P}_{quit}$,\
the mean $P_{quit}$ value over all the users (including unmatched ones).\
After an initial linear growth, we can see that the number of users begins\
to stagnate and slowly decrease.\
Once a sufficient offering to satisfy the majority of users is achieved,\
adding extra \gls{sla} options only increases search difficulty.\
This is seen from the steadily increasing $\overline{P}_{quit}$,\
the probability that a user will quit the search before finding\
a positive-utility \gls{sla}.\
For our scenario, the optimal number of converted customers is achieved between\
6 and 14 \gls{sla}s, depending on the $P_{quit}$ random variable realisations.\
By applying the bootstrap confidence interval method,\ 
we calculate the 95\% confidence interval (CI) for the optimal number of \gls{sla}s\
to be between 8 and 10.


%
%

\vspace{\sectionendmargin}

\section{Conclusion}
\label{sec:conclusion}
\vspace{\sectiontitlemargin}
We presented a novel progressive \gls{sla} specification\
suitable for energy-aware cloud management.\ 
We obtained cloud management traces from two schedulers\ 
optimised for real-time electricity prices\
and temperature-dependent cooling efficiency.\
The \gls{sla}s are derived using\
a method for a posteriori probabilistic modelling of cloud management data\
to estimate upper bounds for \gls{vm} availability, energy savings\
and the resulting \gls{vm} prices.\
The \gls{sla} specification is evaluated\
in a utility-based user \gls{sla} selection simulation\
using realistic workload traces from Wikipedia and Grid'5000.\
Results show mean energy savings per \gls{vm} of up to \ensavings{} due to\
applying more aggressive energy preservation\
actions on users with lower \gls{qos} requirements.\
Furthermore, a wider spectrum of user types with requirements\
not matched by the traditional high availability \gls{vm}s can be reached,\
increasing customer conversion.\

In the future, we plan on expanding the probabilistic model with\
time series forecasting for more accurate \gls{sla} metrics.\
Additional \gls{tc}s could be added to represent other cloud management\
methods, such as the kill-and-restart pattern used on stateless application\
containers in modern web application architectures.\
We also plan to explore \gls{sla} violation detection and how it could\
be integrated into our \gls{sla} specification.\
Furthermore, as predictions change based on day-night and seasonal changes,\
exploring time-changing \gls{sla}s\
in the manner of stocks and bonds\
to match the volatile geotemporal inputs would be feasible. 

\textbf{Acknowledgements.} The work described in this paper has been funded\
through the Haley project (Holistic Energy Efficient Hybrid Clouds)\
as part of the TU Vienna Distinguished Young Scientist Award 2011.

\vspace{\sectionendmargin}

\bibliographystyle{splncs03-kermit}
\bibliography{vmpricing,foued}

\end{document}